\documentclass[twocolumn,natbib]{svjour2}
\bibpunct{[}{]}{,}{n}{}{,} 

\usepackage{graphicx}
\usepackage{amssymb,amsmath}
\usepackage[english]{babel}
\usepackage[utf8]{inputenc}
\usepackage{hyperref}
\usepackage{bm}

\journalname{Few Body Systems}

\title{Three-body recombination rates near a Feshbach resonance within a two-channel contact interaction model}

\author{Peder K. S\o rensen\and Dmitri V. Fedorov \and Aksel S. Jensen}

\institute{P. K. S\o rensen\at
              Aarhus University \\
              \email{pks05@phys.au.dk}
           \and
           D. V. Fedorov \at
              Aarhus University \\
              \email{fedorov@phys.au.dk}
           \and
           A. S. Jensen \at
              Aarhus University \\
              \email{asj@phys.au.dk}
}

\date{} 

\newcommand{\op}{\mathrm{open}} 
\newcommand{\cl}{\mathrm{closed}}

\begin{document}
\maketitle

\begin{abstract}
We calculate the three-body recombination rate into a shallow dimer in a gas of cold bosonic atoms near a Feshbach resonance using a two-channel contact interaction model. The two-channel model naturally des\-cribes the variation of the scattering length through the Feshbach resonance and has a finite effective range. We confront the theory with the available experimental data and show that the two-channel model is able to quantitatively describe the existing data. The finite effective range leads to a reduction of the scaling factor between the recombination minima from the universal value of 22.7. The reduction is larger for larger effective ranges or, correspondingly, for narrower Feshbach resonances.
\end{abstract}

\section{Introduction}\label{intro}
Quantum-mechanical three-body systems of identical bo\-sons exhibit universal features when the two-body scattering length becomes exceedingly large. In this limit---called the universal regime---the properties of the system depend largely on the scattering length alone and can be described by a universal one-channel zero-range model~\cite{Nielsen2001,Braaten2006259}.

The one-channel zero-range model predicts, in particular, that in the limit of large positive scattering length---with a shallow dimer---the low-energy recombination rate of three identical bosons into a shallow dimer exhibits, as function of the scattering length, a geometric scaling: characteristic periodic minima in logarithmic scale with the period equal to~22.7.

Experimentally the recombination rate in the universal regime can be investigated in ultra-cold atomic gases using the technique of Feshbach resonance which enables fine-tuning of the atom-atom scattering length over several orders of magnitude~\cite{chin,Stenger,Kraemer2006}. The characteristic scaling in the recombination rate has been recently observed for some alkali atoms, notably ${}^{39}$K \cite{Zaccanti} and ${}^{7}$Li \cite{Pollack}.

The one-channel models can only describe Feshbach resonances phenomenologically, through a parametrization of the scattering length as function of the external magnetic field. However recently a two-channel model---which provides a natural microscopic description of Fesh\-bach resonances~\cite{bruun2005,nygaard2006}---has been suggested for three-body calculations~\cite{macek2002,macek2002b,PhysRevA.78.020701}.  In addition, the two-channel model has finite effective range---inversely proportional to the width of the Feshbach resonance---and can be expected to describe the deviations from the universal predictions.  This model, however, has not been extensively compared with experiments.

The purpose of this investigation is to confront the two-channel model with the newly available experimental data on the recombination rates, and to determine the effects of the finite effective range on the scaling properties of the recombination rates.

\section{Two-channel contact interaction model}\label{twoch_model}
Feshbach resonance in scattering of cold atoms arises due to an interplay between several coupled channels. The essential physics of the resonance can be described by a model with two coupled channels~\cite{bruun2005,nygaard2006}.

The system of two identical bosonic atoms in an $s$-wave state is described by a two component wave-function $\psi$. In the following we assume that the energy of the system is below the excitation threshold and refer to the ground state and the excited state channels as correspondingly open and closed,

\begin{equation}
	\psi(r)=\frac1r\begin{bmatrix}u_\cl(r)\\ u_\op(r)\end{bmatrix},
	\label{eq:1}
\end{equation}
where $r$ is the distance between atoms, and the radial functions $u_\op$ and $u_\cl$ describe correspondingly the open channel where the two atoms are in the ground state, and the closed channel where one of the atoms is excited.

In a contact interaction model the radial components of the wave-function satisfy two free Schrödinger equations,
\begin{subequations}\label{eq:2}
\begin{align}
	-\frac{\hbar^2}{2m^*}u_\cl'' &= (E-E^*)u_\cl \;,\label{eq:2a}\\
	-\frac{\hbar^2}{2m^*}u_\op'' &= Eu_\op\;,
	\label{eq:2b}
\end{align}
\end{subequations}
where primes denote differentiation with respect to the relative distance $r$, $m^*$ is the reduced mass of the two atoms, $E$ is the energy of the system, and $E^*$ is the excitation energy from the ground to the excited state of the atom.

The interaction between atoms appear in the contact interaction model only through a non-trivial boundary condition at small separation between atoms,
\begin{equation}
	\begin{bmatrix}u_\cl'\\u_\op'\end{bmatrix}_{r=0} =
	\begin{bmatrix}-a_\cl^{-1}&\beta\\\beta&-a_\op^{-1}\end{bmatrix}
	\begin{bmatrix}u_\cl\\u_\op\end{bmatrix}_{r=0} \;,
	\label{eq:3}
\end{equation}
where the constant $\beta$ parametrises the coupling between the channels, and $a_\op$ and $a_\cl$ are the interaction parameters which become scattering lengths in the corresponding channels in the limit of vanishing coupling. The model is similar to that of \cite{macek2002b} but is more general since we allow $a_\cl\neq a_\op$.

For scattering below threshold, $0<E<E^*$, the solutions to the Schrödinger equation \eqref{eq:2} should be sought in the form
\begin{subequations}\label{eq:4}
\begin{align}
	u_\cl &= A_\cl e^{-\kappa_\cl r} \;,\\
	u_\op &= A_\op\sin(k_\op r+\delta) \;,
\end{align}
\end{subequations}
where $A_\cl$ and $A_\op$ are constants, $k_\op=\sqrt{2m^*E/\hbar^2}$, and
$\kappa_\cl=\sqrt{2m^*(E^*-E)/\hbar^2}$.

Inserting ansatz \eqref{eq:4} into the boundary condition
\eqref{eq:3} gives
\begin{equation}
	\begin{bmatrix}-\beta\sin\delta & a_\cl^{-1}-\kappa_\cl\\
	k_\op\cos\delta+a_\op^{-1}\sin\delta & -\beta\end{bmatrix}
	\begin{bmatrix}A_\cl\\A_\op\end{bmatrix}=0\;.
	\label{eq:5}
\end{equation}

The homogeneous linear system \eqref{eq:5} has non-trivial solutions only when the determinant of the matrix vanishes
\begin{equation}
	\left(k_\op\cos\delta+\frac{\sin\delta}{a_\op}\right)
	\left(\frac1{a_\cl}-\kappa_\cl\right)-\beta^2\sin\delta=0 \;.
	\label{6}
\end{equation}
Solving this equation gives the scattering phase,
\begin{equation}
	k_\op\cot\delta=-\frac{1}{a_\op}+\frac{\beta^2}{a_\cl^{-1}-\kappa_\cl} \;.
	\label{eq:7}
\end{equation}
Expanding~\eqref{eq:7} for small $k_\op$ using $\kappa_\cl^2=\kappa^2-k_\op^2$, where $\kappa^2=2m^*E^*/\hbar^2$,
gives the effective range expansion,
\begin{equation}
	k_\op\cot\delta = -\frac{1}{a}+\frac12 R k_\op^2 +O(k_\op^4) \;,
	\label{eq:8}
\end{equation}
where $a$ and $R$ are the scattering length and effective range respectively,
\begin{align}
	\frac1a&=\frac{1}{a_\op}+\frac{\beta^2}{\kappa-a_\cl^{-1}} \;,\label{eq:9} \\
	R&=-\frac{\beta^2}
	{\kappa\left(\kappa-a_\cl^{-1}\right)^2} \;.
	\label{eq:10}
\end{align}
The two-channel contact interaction model thus has a negative finite effective range.

\section{Feshbach resonance and model parameters} 
If an external magnetic field is applied to the system, the excitation energy $E^*$ is substituted with
\begin{equation}
	E^*\rightarrow E^*-~\delta\mu B\;,
	\label{eq:}
\end{equation}
where $B$ is the applied magnetic field, and $\delta\mu$ is the difference between the magnetic moments of the atom in the ground and the excited state. The scattering length \eqref{eq:9} is then a function of the magnetic field,
\begin{equation}
	a(B)=a_\op\frac{\kappa(B)-a_\cl^{-1}}{\kappa(B)-a_\cl^{-1}+\beta^2a_\op} \;,
	\label{eq:11}
\end{equation}
where $\kappa(B)=\sqrt{2m^*(E^*-\delta\mu B)/\hbar^2}$.

The scattering length diverges at a critical value of the magnetic field, $B_0$, given by 
\begin{equation}
	\kappa(B_0)=\kappa_0=\frac1{a_\cl}-\beta^2a_\op\;,
	\label{eq:12}
\end{equation}
which gives
\begin{equation}
	B_0 = \frac1{\delta\mu} \left( E^* - \frac{\hbar^2\kappa_0^2}{2m^*} \right)\;.
	\label{eq:13}
\end{equation}
Expanding $a(B)$ in the vicinity of $B_0$ gives
\begin{equation}
	a(B) \approx a_\op\left(1 - \frac{\Delta B}{B-B_0} \right) \;,
	\label{eq:14}
\end{equation}
where 
\begin{equation}
	\Delta B = \frac1{\delta\mu} \frac{\hbar^2\kappa_0\beta^2a_\op}{m^*}\;.
	\label{eq:15}
\end{equation}

The characteristic dependence~\eqref{eq:14} of the scattering length on the magnetic field is referred to as Feshbach resonance~\cite{chin} and is well known empirically. The two-channel contact interaction model is thus able to naturally describe this phenomenon.

In the vicinity of the Feshbach resonance the effective range is inversely proportional to the width $\Delta B$ of the resonance. Indeed from \eqref{eq:9} with $a=\infty$ we get
\begin{equation}
	R(B_0) = -\frac1{\kappa_0\beta^2a_\op^2}= -\frac1{a_\op}\frac{\hbar^2}{m^*\delta\mu\Delta B}\;,
	\label{eq:16}
\end{equation}
in agreement\footnote{Note that $m^*=m/2$ where $m$ is the mass of the atom.} with \cite{bruun2005}. This value holds precisely at the resonance only. The $B$-dependence of $R$ is, from \eqref{eq:10}
\begin{equation}
	R(B)=R(B_0)\left(1-\frac{a_\op}{a(B)}\right)^2\;.
	\label{eq:16a}
\end{equation}

\begin{figure}[h]
\centering
\input{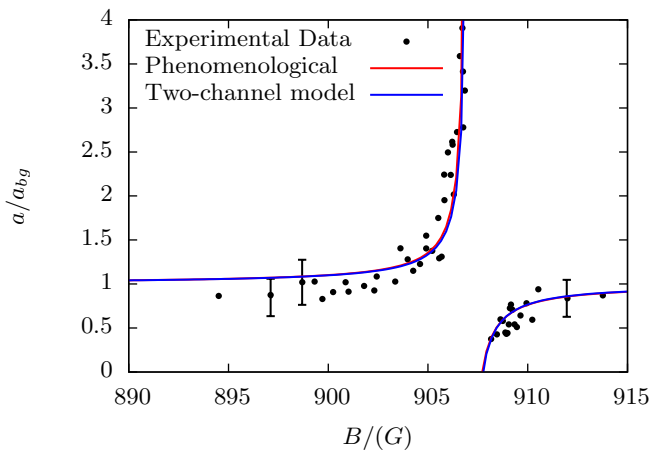}
\caption{Scattering length $a$ between two ${}^{23}$Na atoms as a function of external magnetic field strength $B$ with two-channel \eqref{eq:9} and empirical \eqref{eq:14} fits to experimental data from \cite{Stenger}. The Feshbach resonance is seen at $B_0=907$ G with width $\Delta B=0.7$ G.}
\label{Na23Feshbach}
\end{figure}

Given $\Delta B,\;B_0$ and the background scattering length $a_{bg}$ from experiment, the equations \eqref{eq:9}, \eqref{eq:13} and \eqref{eq:15} can be solved for the model parameters $a_\op,\;a_\cl$ and $\beta$,
\begin{subequations}\label{eq:17}
\begin{align}
	a_\op&=a_{bg}\;,\label{eq:17a}\\
	a_\cl&=\frac{\textrm{sign}(\Delta B)2\sqrt\epsilon}{\frac{\delta\mu\Delta B}{E_0}+2\epsilon}a_\op\;,\label{eq:17b}\\
	\beta^2&=\frac{1}{2a_\op^2}\frac{1}{\sqrt \epsilon}\frac{\delta\mu|\Delta B|}{E_0}\;,
	\label{eq:17c}
\end{align}
\end{subequations}
where
\begin{equation}
	\epsilon=\frac{E^*-\delta\mu B_0}{E_0},\quad E_0=\frac{\hbar^2}{2m^*a_{bg}^2}\;.
	\label{eq:18}
\end{equation}
The value of $E^*$ cannot be determined uniquely from the equations. It can, however, be found by fitting \eqref{eq:9} to experimental data points $a(B)$ as done on figure \ref{Na23Feshbach}. However, the value of $E^*$ does not affect the final observables significantly, provided it is greater than $\delta\mu B_0$ and is of the order the hyperfine splitting.

Figure \ref{Na23Feshbach} shows experimental data for a Feshbach reso\-nance in ${}^{23}$Na along with fits \eqref{eq:9} and \eqref{eq:14}. The phenomenological expression \eqref{eq:14} gives $B_0=907.0$ G, $\Delta B=0.71$ G, while the two-channel expression \eqref{eq:9} gives $B_0=907.1$ G, $\Delta B=0.69$ G when using the expressions \eqref{eq:13} and \eqref{eq:15}. Choosing $E^*=23\;\mu$eV provides a good fit. Varying $E^*$ has relatively little influence on the values of $B_0$ and $\Delta B$ provided it lies within this order of magnitude.

The phenomenological and the two-channel curves are virtually identical, so in the following we use the experimental parameters $a_{bg},\;\Delta B,\;B_0$ and $\delta\mu$ and determine the parameters of the two-channel model through the equations \eqref{eq:17}.

\section{Three-body hyperspherical adiabatic method}\label{hyperspheric_adiabatic}
We describe the system of three particles using hyper\-spherical coordinates defined from the Cartesian coordinates $\bm r_i$,$\bm r_j$,$\bm r_k$, of particles $i,j,k$ as \cite{FedorovJensen2001}
\begin{gather}
	\bm x_i=\sqrt{\mu_i}(\bm r_j-\bm r_k),\\\bm y_i=\sqrt{\mu_{jk}}\left(\bm r_i-\frac{m_j\bm r_j+m_k\bm r_k}{m_j+m_j}\right),\\
	\mu_i=\frac1m\frac{m_jm_k}{m_j+m_k},\qquad \mu_{jk}=\frac1m\frac{m_i(m_j+m_k)}{m_i+m_j+m_k},\label{eq:20}\\
	\rho^2=x_i^2+y_i^2,\qquad\rho\sin\alpha_i=x_i,\qquad\rho\cos\alpha_i=y_i,
	\label{eq:21}
\end{gather}
where $\{i,j,k\}$ are cyclic permutations of $\{1,2,3\}$, $\rho$ is the hyperradius and $\alpha$ is one of the hyperangles, the four remaining hyperangles determine the direction of $\bm x_i$ and $\bm y_i$. All five hyperangles are denoted collectively as $\Omega$. The constant $m$ is a mass-scaling parameter that we choose to be the mass of the atoms such that $\mu_i=\frac{1}{2}$ for all $i$. The index $i$ is referred to as the Jacobi-index of the chosen coordinate set.

We expand the wave-function $\Psi$ on adiabatic basis states $\Phi_n(\rho,\Omega)$,
\begin{equation}
	\Psi(\rho,\Omega)=\rho^{-5/2}\sum_nf_n(\rho)\Phi_n(\rho,\Omega),
	\label{eq:22}
\end{equation}
where $\Phi_n(\rho,\Omega)$, are solutions to the hyperangular equation \cite{Nielsen2001}
\begin{equation}
	\left(\Lambda+\frac{2m\rho^2}{\hbar^2}V\right)\Phi_n(\rho,\Omega)=\lambda_n(\rho)\Phi_n(\rho,\Omega),
	\label{eq:23}
\end{equation}
where $\Lambda$ is the grand angular momentum operator in hyperradial coordinates and $\lambda_n$ is the eigenvalue, the index $n$ refers to the adiabatic channel. The hyper-radial functions $f_n(\rho)$ satisfy the hyper-radial equations
\begin{multline}
\left(-\frac{d^2}{d\rho^2}+\frac{\lambda_n+15/4}{\rho^2}-Q_{nn}(\rho)-\frac{2mE}{\hbar^2}\right)f_n(\rho)\\
=\sum_{m\neq n}\left(2P_{nm}(\rho)\frac{d}{d\rho}+Q_{nm}(\rho)\right)f_{m}(\rho)
	\label{eq:4.7}
\end{multline}
where the adiabatic coupling terms $P$ and $Q$ are defined by
\begin{equation}
P_{nm}(\rho) = \langle\Phi_n|\frac{\partial}{\partial\rho}|\Phi_{m}\rangle\,,\qquad Q_{nm}(\rho) = \langle\Phi_n|\frac{\partial^2}{\partial\rho^2}|\Phi_{m}\rangle\,,
	\label{eq:4.8}
\end{equation}
with brackets indicating integration over all hyper-angles.

Contact interaction potentials are equal to zero for $\rho>0$, and are specified by applying boundary conditions on $\Phi_n$. The boundary condition \eqref{eq:3} in hyperspherical coordinates becomes
\begin{equation}
	\frac{\partial(\alpha_i\Phi)}{\partial\alpha_i}\bigg|_{\alpha_i=0}=\frac{\rho}{\sqrt{\mu_i}}\begin{bmatrix}-a_{i,\cl}^{-1}&\beta_i\\\beta_i&-a_{i,\cl}^{-1}\end{bmatrix}\alpha_i\Phi\bigg|_{\alpha_i=0},
	\label{eq:24}
\end{equation}
where $a_{i,\op}$, $a_{i,\cl}$ are scattering lengths for, and $\beta_i$ coupling strength between the open and closed channel of particles $j$ and $k$, and $\Phi=\left[\begin{smallmatrix}\Phi_\cl\\\Phi_\op\end{smallmatrix}\right]$ is a two-component wave-function that describes the two interaction channels. Since the particles are identical, in the following we choose a specific Jacobi-set and suppress the indices.

We use Faddeev decomposition of the hyperangular wave-function with $s$-wave states only, $\Phi=\phi_1+\phi_2+\phi_3$, with 
\begin{equation}
	\phi_i=\begin{bmatrix}\tilde N\sin\left(\tilde\nu\left[\alpha_i-\frac{\pi}{2}\right]\right)\\N\sin\left(\nu\left[\alpha_i-\frac{\pi}{2}\right]\right)\end{bmatrix}\;,
	\label{eq:25}
\end{equation}
where $N$ and $\tilde N$ are amplitudes and where
\begin{equation}
	\nu^2=\lambda+4
	\label{eq:25a}
\end{equation}
and 
\begin{equation}
	\tilde\nu^2=\nu^2-\kappa^2\rho^2\;.
	\label{eq:25b}
\end{equation}
The solutions $\phi_j$ and $\phi_k$ must be expressed in the same coordinate system as $\phi_i$ in order to use the same boundary condition. This is done by a rotation and projection operator $\mathcal R$ \cite{FedorovJensen2001}
\begin{equation}
	\mathcal R[\phi_k]=\frac1{\sin(2\varphi_{ik})}\int_{|\varphi_{ik}-\alpha_i|}^{\frac\pi2-|\frac\pi2-\varphi_{ik}-\alpha_i|}\phi_{k}(\alpha_k)d\alpha_k
	\label{eq:4.16}
\end{equation}
where $\varphi_{ik}=\frac{\pi}{6}$ for identical particles. This yields
\begin{equation}
	\Phi(\alpha_i)=\phi(\alpha_i)+2\mathcal R[\phi_k](\alpha_i)\;.
	\label{eq:4.16a}
\end{equation}
Plugging this solution into the boundary condition~\eqref{eq:24} gives a matrix equation similar to \eqref{eq:5} of the two-particle model. Again a solution exists only when the determinant equals zero, yielding the eigenvalue equation
\begin{equation}
	\frac{\rho^2\beta^2}{\mu}\sin\left(\nu\frac\pi2\right)\sin\left(\tilde\nu\frac\pi2\right)-f_\op(\nu)f_\cl(\tilde\nu)=0,
	\label{eq:26}
\end{equation}
where $\kappa$ is related to the energy sepa\-ration $E^*$ between the channels by $E^*=\hbar^2\kappa^2/2m^*$, and the function $f_l$ is (with $l=\mathrm{open},\mathrm{closed}$)
\begin{equation}
	f_l(x)=x\cos\left(x\frac\pi2\right)-\frac8{\sqrt3}\sin\left(x\frac\pi6\right)-\frac\rho{\sqrt\mu}\frac1{a_l}\sin\left(x\frac\pi2\right)\;,
	\label{eq:27}
\end{equation}
where $x=\nu$ when $l=\mathrm{open}$ and $x=\tilde\nu$ when $l=\mathrm{closed}$. This implicitly defines the eigenvalue $\nu(\rho)$ which is the central quantity from which all other results follow. 

\section{WKB method for recombination rate}\label{hidden_crossing_wkb}
\begin{figure}
\centering
\input{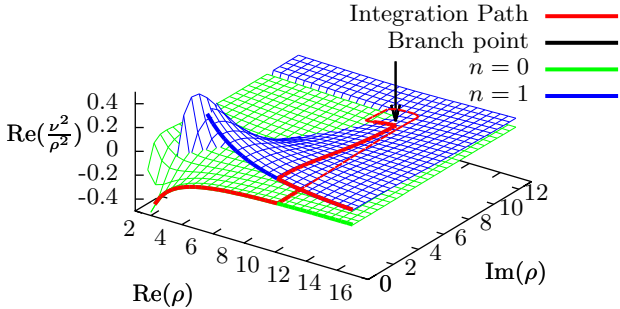}
\caption{Illustration of the integration path taken in the hidden crossing method.}
\label{crossing}
\end{figure}

The recombination rate is given by \cite{Zaccanti}
\begin{equation}
	\dot n=-\alpha n^3\;,
	\label{eq:28}
\end{equation}
where $n$ is the particle density and $\alpha$ is the recombination coefficient. We calculate the recombination coefficient using the WKB method of hidden crossing theory \cite{Macek1999} where $\alpha$ is given as
\begin{equation}
	\alpha=8(2\pi)^23\sqrt3\frac\hbar{m^*}\lim_{k\rightarrow0}\frac{P(k)}{k^4},
	\label{eq:29}
\end{equation}
with the wave number $k$ defined by $E=\hbar^2k^2/2m^*$. The transition probability $P(k)$ is
\begin{equation}
	P(k)=4e^{-2S}\sin^2\Delta\;,
	\label{eq:30}
\end{equation}
where
\begin{equation}
	\Delta+iS=\int_cd\rho\sqrt{k^2-\frac{\nu(\rho)^2}{\rho^2}}\;,
	\label{eq:31}
\end{equation}
and the integral is taken along a contour in the complex $\rho$-plane connecting the adiabatic channel $n=1$ corresponding to three free particles to the channel $n=0$ corresponding to a shallow dimer and a free particle. Figure \ref{crossing} is a visualization of the contour along which the integral is calculated. Note that this expression has included the Langer correction term  $1/4\rho^2$ \cite{PhysRev.51.669} which in this context modifies the effective radial potential
\begin{equation}
	\frac{\lambda+15/4}{\rho^2}+\frac{1}{4\rho^2}=\frac{\nu^2}{\rho^2}
	\label{eq:31a}
\end{equation}
The $Q_{nn}$ terms are not included in hidden crossing theory \cite{PhysRevA.54.544}.

The integration path must enclose a branch-point $\rho_b$ in order to connect channels. The branch-point is found by solving \cite{Nielsen2001}
\begin{equation}
	\frac{d\rho}{d\nu}\bigg\vert_{\nu_\text b}=0,
	\label{eq:32}
\end{equation}
for (complex) $\nu_b$ and evaluating $\rho_b=\rho(\nu_b)$. In the one-channel model the branch-point is at $\rho_b\approx(2.592 + 2.974i)\sqrt\mu a$. For the two-channel model $\rho(\nu)$ is given only implicitly and \eqref{eq:32} must be solved numerically. The above value for $\rho_b$ is, however, approximately correct.

Furthermore, in the two-channel model there exist a second branch-point between the channel $n=0$ and $n=1$ since the eigenvalue equation \eqref{eq:26} is second order in $\nu$. This means that the transition probability \eqref{eq:30} is modified. One has to take the coherent sum of the contributions from the two paths 
\begin{equation}
	P(k)=\left|2e^{-S_1}\sin\Delta_1+2e^{-S_2}\sin\Delta_2\right|^2,
	\label{eq:33}
\end{equation}
where the indices denote the path taken. However, the scattering length has to be rather small compared to the effective range for this additional contribution to the recombination coefficient to be noticeable.



\section{Thomas effect and recombination rate scaling}

In the one-channel model the angular eigenvalue equation is given
as \cite{FedorovJensen2001,efimov1970}
\begin{equation}
  \nu \cos \left( \nu \frac{\pi}{2} \right) - \frac{8}{\sqrt{3}} \sin \left(
  \nu \frac{\pi}{6} \right) = \frac{1}{\sqrt{\mu}} \frac{\rho}{a} \sin \left(
  \nu \frac{\pi}{6} \right), \label{eq:oc}
\end{equation}
which can be easily obtained from \eqref{eq:26} by setting $\beta = 0$. At small
distances, $\rho \ll a$, this equation has a specific imaginary solution
\begin{equation}
  \nu_s \approx 1.006 i \;.
\end{equation}
This solution creates an exceedingly attractive hyper-radial potential at
small distances,
\begin{equation}
  \frac{\nu_s^2 - \frac{1}{4}}{\rho^2} \approx - \frac{1.262}{\rho^2}.
\end{equation}
In the lowest hyper-radial equation \eqref{eq:4.7} this leads to the collapse of the system, known as the Thomas effect \cite{thomas,PhysRevA.63.063608}.

It has been shown \cite{FedorovJensen2001,PhysRevA.80.013608} that incorporating a finite effective
range in the single-channel approach removes the Thomas effect. Indeed,
introducing the effective range correction in the boundary condition \cite{FedorovJensen2001} leads to
the single-channel eigenvalue equation
\begin{multline}
  \nu \cos \left( \nu \frac{\pi}{2} \right) - \frac{8}{\sqrt{3}} \sin \left(
  \nu \frac{\pi}{6} \right) =\\ \frac{\rho}{\sqrt{\mu}} \left[ \frac{1}{a} -
  \frac{1}{2} R \left( \frac{\sqrt{\mu} \nu}{\rho} \right)^2 \right] \sin
  \left( \nu \frac{\pi}{6} \right),
\end{multline}
where that imaginary solution---and correspondingly the Thomas
collapse---is removed at small distances.

However, in the two-channel contact interaction model ---despite its
finite effective range---the Thomas collapse persists. Indeed, at $\rho
= 0$ the eigenvalue equation \eqref{eq:27} turns into the square of the (corresponding
limit of) the single-channel equation (\ref{eq:oc}),
\begin{equation}
  \left[ \nu \cos \left( \nu \frac{\pi}{2} \right) - \frac{8}{\sqrt{3}} \sin
  \left( \nu \frac{\pi}{6} \right) \right]^2 = 0,
\end{equation}
where the imaginary root $\nu_s$ and hence the Thomas effect are still
present. Therefore the two-channel model---just like the unmodified
one-channel model---needs a cut-off at small distances.

This imaginary root in the one-channel model leads to a characteristic scaling
of the recombination rate as function of the scattering length $a$ \cite{Nielsen2001,Macek1999}. Indeed, since $a$ is the only parameter of
the model, the real part of the WKB integral \eqref{eq:31} can, in the large $a$ and low energy limit, be estimated as
\begin{equation}
  \Delta \cong \int_{\rho_c}^a d \rho \frac{\left| \nu_s \right|}{\rho} =
  \left| \nu_s \right| \ln \left( \frac{a}{\rho_c} \right),
  \label{deltaapprox}
\end{equation}
where $\rho_c$ is the regularization cut-off. The recombination
rate---having $\sin^2 \Delta$ as a factor---will then exhibit, as
function of $a$, a characteristic geometric series of minima at the points
determined by the zeros of the sine function,
\begin{equation}
  \left| \nu_s \right| \ln \left( \frac{a}{\rho_c} \right) = \pi n \quad,\quad n = 1, 2, \ldots .
\end{equation}
The periodic factor of the series is equal
\begin{equation}
  \exp \left( \frac{\pi}{\left| \nu_s \right|} \right) \approx 22.7
\end{equation}
In the two-channel model the upper limit in the estimate \eqref{deltaapprox} should be
modified due to the existence of another length parameter, the effective range.
Correspondingly, the scaling law is also expected to be modified.

\section{Comparison with experimental data}\label{exp_data}
We compare the two-channel model with the experimental data for cold atomic gasses listed in table \ref{tab:data}. The effective ranges are calculated using the formula \eqref{eq:16}.
\setlength{\tabcolsep}{4.5pt}
\begin{table}[h]
	\caption{Experimental data for Feshbach resonances for three atomic gasses. $\mu_B$ is the Bohr magneton and $a_0$ the Bohr radius.}
	\centering
	\label{tab:data}
	\begin{tabular}{rccccc}
		\hline\noalign{\smallskip}
		 & $B_0$ [G] & $\Delta B$ [G] & $\delta\mu$\,[$\mu_\text B$] & $a_{bg}\,[a_0]$ & $R_\text{eff}\,[|a_{bg}|]$\\[3pt]
		\tableheadseprule\noalign{\smallskip}
		${}^{23}$Na \cite{Stenger} & 907 & 0.70 & 3.8 & 63 & -21\\
		${}^{133}$Cs \cite{Kraemer2006} & -11.7 & 28.7 & 2.3 & 1720 & -1.99$\times10^{-4}$\\
		${}^{39}$K \cite{Zaccanti} & 402.4 & -52 & 1.5 & -29 & -2.02\\
		${}^{7}$Li \cite{Pollack} & 736.8 & -192.3 & 1.93 & -25 & -3.17\\
		\noalign{\smallskip}\hline
	\end{tabular}
\end{table}

In figure \ref{Na23} the result from the two-channel model is shown together with the experimental data for ${}^{23}$Na. The cut-off is fixed by the experimental minimum at $a_1^*=62a_0$. The rather large effective range could make the finite range effect ---reduction of the scaling factor down to 15.7--- noticeable. However, at least one additional minimum is needed to make a proper comparison. Experimental data is not yet available for this range.

\begin{figure}
\centering
\input{Na23}
\caption{Recombination coefficient $\alpha$, eq. \eqref{eq:29}, for ${}^{23}$Na as a function of scattering length $a$ from the two-channel model compared with the experimental data from \cite{Stenger}. The theory predicts the next minimum to be around $a_2^*\approx1000a_0$.}
\label{Na23}
\end{figure}

In figure \ref{Cs133} the result from the two-channel model is shown together with the experimental data for ${}^{133}$Cs. The cut-off is fixed by the experimental minimum at $a_1^*\approx210a_0$. The effective range is very small indeed and the results from the two-channel model are virtually indistinguishable from the one-channel model with the scaling factor of 22.7. The next minimum should be found at $a_2^*\approx4770a_0$.

\begin{figure}
\centering
\input{Cs133}
\caption{Recombination coefficient $\alpha$, eq. \eqref{eq:29}, for ${}^{133}$Cs as a function of scattering length $a$ from the two-channel model compared with the experimental data from \cite{Kraemer2006}. The theory predicts the next minimum to be around $a_2^*\approx4770a_0$.}
\label{Cs133}
\end{figure}

\begin{figure}
\centering
\input{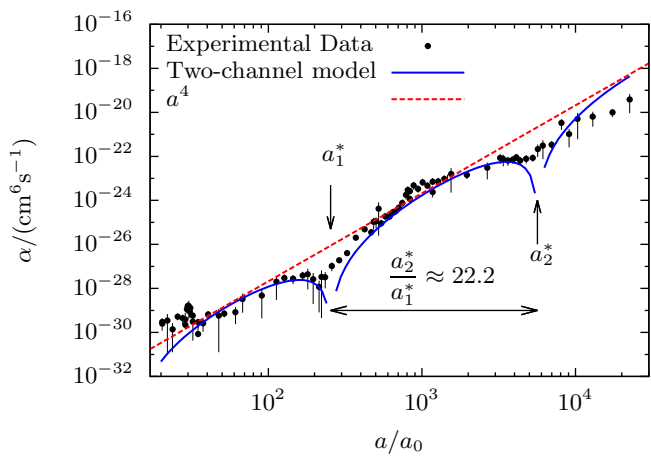}
\caption{Recombination coefficient $\alpha$, eq \eqref{eq:29}, for ${}^{39}$K as a function of scattering length $a$ from the two-channel model compared with the experimental data from \cite{Zaccanti}. The overall $a^4$ scaling is also shown.}
\label{K39}
\end{figure}

Figure \ref{K39} shows the recombination coefficient for ${}^{39}$K. The recombination minimum $a_2^*=5650\pm900$ is chosen to fit the cut-off parameter. The two-channel model gives $a_1^*=254a_0$, with the experimental value of $a_1^*=(224\pm7)a_0$. Overall the two-channel model fits the data quite well. Notably the scaling is correct compared to experiment. The ratio of minima from the two-channel model is 22.2 whereas the experimental value is $25.2\pm4.1$. Our result thus lies within the experimental uncertainty.

\begin{figure}
\centering
\input{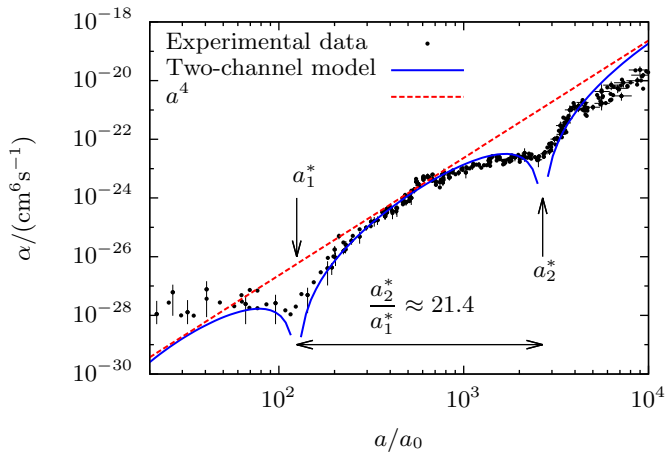}
\caption{Recombination coefficient $\alpha$, eq. \eqref{eq:29}, for ${}^{7}$Li as a function of scattering length $a$ from the two-channel model compared with the experimental data from \cite{Pollack}. The overall $a^4$ scaling is also shown.}
\label{Li7}
\end{figure}

The recombination coefficient for ${}^7$Li is shown in figure \ref{Li7}. The two minima are at $a_1^*=(119\pm11)a_0$ and $a_2^*=(2676\pm195)a_0$. The cut-off is fixed by $a_2^*$, giving the two-channel prediction $a_1^*=125a_0$. Again the theory describes the experimental data very well. The two-channel model ratio of minima is 21.4 while the experimental value is $22.5\pm2.6$ and again our result lies within the experimental uncertainty.

At small $a$ the experimental recombination coefficient is two orders of magnitude higher than the models predict (a similar tendency is seen in figure \ref{K39}). This is possibly due to finite range of the physical potentials which cannot be correctly modelled by the presented contact interaction potentials.

\begin{figure}
\centering
\input{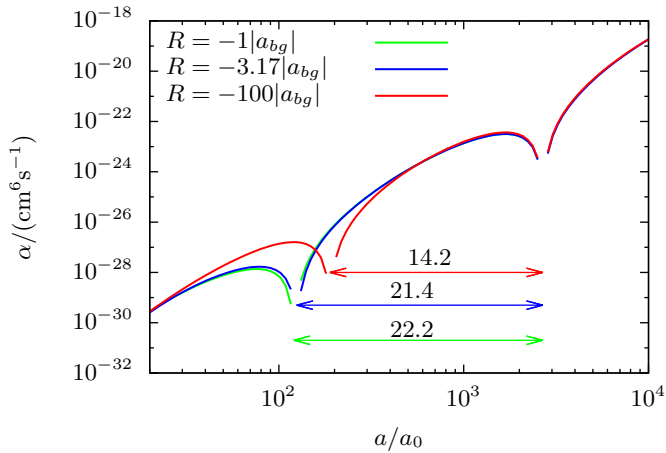}
\caption{Recombination coefficient $\alpha$, eq. \eqref{eq:29}, for ${}^{7}$Li as a function of scattering length $a$ from the two-channel model with different effective ranges as an illustration of their effect on the recombination minima. The scaling factor of ratios of minima is reduced the large $|R|$ becomes.}
\label{REffects}
\end{figure}

\section{Finite range effects}
As illustrated on the figures \ref{Na23}, \ref{K39}, and \ref{Li7} the two-channel model---with its finite effective range---shows a reduction of the scaling factor between the recombination minima as compared to the zero-range model value of 22.7 -- down to 15.7 for ${}^{23}$Na, 22.2 for ${}^{39}$K, and 21.4 for ${}^{7}$Li.

A similar reduction from the zero-range result due to finite-range effects was also observed in the bound state spectrum \cite{PhysRevA.78.020501}.

For ${}^{133}$Cs there was no noticeable change due to the very small effective range. The relatively small reduction in the cases of ${}^{39}$K and ${}^{7}$Li is apparently due to the small effective ranges, which in turn is due to the relatively large widths of the corresponding Feshbach resonances, see table~1.

In the case of ${}^{23}$Na the effective range is an order of magnitude larger and, correspondingly, the reduction of the factor is also larger. Unfortunately at the moment there are no experimental data around the second minimum to confirm the prediction of the two-channel model concerning its position of the second minimum.

On figure \ref{REffects} we illustrate the effects of the effective range on the recombination rate by performing exploratory calculations of the rate for ${}^7$Li where we vary the effective range---by varying $\Delta B$---keeping the other parameters unchanged. Indeed increasing the effective range decreases the scaling factor between the minima. However, for a noticeable effect the effective range has to be of the order of several dozen $a_{bg}$ which corresponds to the resonance width of less than $10$ G.

\section{Conclusion}\label{conclusion}
We have calculated three-body recombination rates near Feshbach resonances using the two-channel contact interaction model and compared the results with the available experimental data. We have shown that the two-channel model is able to quantitatively describe all available data.

Unlike the ubiquitous one-channel zero-range model the two-channel model naturally describes the magnetic field dependence of the scattering length through a Fesh\-bach resonance. In addition it has finite effective range. However, despite the finite effective range the Thomas effect persists in the two-channel model and a regularization is still needed.

The finite effective range---which in this model is inversely proportional to the width of the resonance---leads to a reduction of the scaling between recombination minima as compared to the zero-range scaling factor of $\approx 22.7$. The effect is, however, relatively small in the investigated datasets for ${}^7$Li, ${}^{39}$K and especially for ${}^{133}$Cs because of the relatively small effective ranges or correspondingly relatively large widths of the corresponding Feshbach resonances.

For ${}^{23}$Na data the Feshbach resonance is an order of magnitude narrower, and the effective range, corres\-pondigly, an order of magnitude larger. The two-channel model then predicts the scaling factor of 15.4. However, this prediction has to be verified experimentally as the data in the vicinity of the second minimum is not yet available.

\bibliographystyle{unsrt}
\bibliography{bibliotek}

\begin{thebibliography}{10}

\bibitem{Nielsen2001}
E.~Nielsen, D.~V. Fedorov, A.~S. Jensen, and E.~Garrido.
\newblock The three-body problem with short-range interactions.
\newblock {\em Physics Reports}, 347(5):373 -- 459, 2001.

\bibitem{Braaten2006259}
E.~Braaten and H.-W. Hammer.
\newblock Universality in few-body systems with large scattering length.
\newblock {\em Physics Reports}, 428(5-6):259 -- 390, 2006.

\bibitem{chin}
C.~Chin, R.~Grimm, P.~Julienne, and E.~Tiesinga.
\newblock Fesh\-bach resonances in ultracold gases.
\newblock {\em Rev. Mod. Phys.}, 82(1225), 2010.

\bibitem{Stenger}
J.~Stenger, S.~Inouye, M.~R. Andrews, H.-J. Miesner, D.~M. Stamper-Kurn, and
  W.~Ketterle.
\newblock Strongly enhanced inelastic collisions in a bose-einstein condensate
  near feshbach resonances.
\newblock {\em Phys. Rev. Lett.}, 82(12), Mar 1999.

\bibitem{Kraemer2006}
T.~Kraemer et~al.
\newblock Evidence for efimov quantum states in an ultracold gas of caesium
  atoms.
\newblock {\em Nature}, 440:315--318, 2006.

\bibitem{Zaccanti}
M.~Zaccanti, B.~Deissler, C.~D'Errico, M.~Fattori, M.~Jona-Lasinio, S.~Muller,
  G.~Roati, M.~Inguscio, and G.~Modugno.
\newblock Observation of an efimov spectrum in an atomic system.
\newblock {\em Nature Physics}, 5, 2009.

\bibitem{Pollack}
S.~E. Pollack, D.~Dries, and R.~G. Hullet.
\newblock Universality in three- and four-body bound states of ultracold atoms.
\newblock {\em Science}, 326, 2009.

\bibitem{bruun2005}
G.~M. Bruun, A.~D. Jackson, and E.~E. Kolomeitsev.
\newblock Multichannel scattering and feshbach resonances: Effective theory,
  phenomenology, and many-body effects.
\newblock {\em Phys. Rev. A.}, 71(052713), 2005.

\bibitem{nygaard2006}
N.~Nygaard, B.~I. Schneider, and P.~S. Julienne.
\newblock Two-channel r-matrix analysis of magnetic-field-induced feshbach
  resonances.
\newblock {\em Phys. Rev. A.}, 73(042705), 2006.

\bibitem{macek2002}
O.~I. Kartavtsev and J.~H. Macek.
\newblock Low-energy three-body recombination near a feshbach resonance.
\newblock {\em Few-Body Systems}, 31:249--254, 2002.

\bibitem{macek2002b}
J.~H. Macek.
\newblock Multichannel zero-range potentials in the hyperspherical theory of
  three-body dynamics.
\newblock {\em Few-Body Systems}, 31:241--248, 2002.

\bibitem{PhysRevA.78.020701}
N.~P. Mehta, S.~T. Rittenhouse, J.~P. D'Incao, and C.~H. Greene.
\newblock Efimov states embedded in the three-body continuum.
\newblock {\em Phys. Rev. A}, 78:020701, 2008.

\bibitem{FedorovJensen2001}
D.~V. Fedorov and A.~S. Jensen.
\newblock Regularization of a three-body problem with zero-range potentials.
\newblock {\em Journal of Physics A}, 34(30), 2001.

\bibitem{Macek1999}
E.~Nielsen and J.~H. Macek.
\newblock Low-energy recombination of identical bosons by three-body
  collisions.
\newblock {\em Phys. Rev. Lett.}, 83(8), Aug 1999.

\bibitem{PhysRev.51.669}
Rudolph~E. Langer.
\newblock On the connection formulas and the solutions of the wave equation.
\newblock {\em Phys. Rev.}, 51:669--676, Apr 1937.

\bibitem{PhysRevA.54.544}
J.~H. Macek and S.~Yu. Ovchinnikov.
\newblock Hyperspherical theory of three-particle fragmentation and wannier's
  threshold law.
\newblock {\em Phys. Rev. A}, 54:544--560, Jul 1996.

\bibitem{efimov1970}
V.~Efimov.
\newblock Energy levels arising from resonant two-body forces in a three-body
  system.
\newblock {\em Phys. Lett. B}, 33, 1970.

\bibitem{thomas}
L.~H. Thomas.
\newblock The interaction between a neutron and a proton and the structure of
  ${\mathrm{h}}^{3}$.
\newblock {\em Phys. Rev.}, 47:903--909, Jun 1935.

\bibitem{PhysRevA.63.063608}
D.~V. Fedorov and A.~S. Jensen.
\newblock Correlation-induced collapse of many-body systems with zero-range
  potentials.
\newblock {\em Phys. Rev. A}, 63:063608, May 2001.

\bibitem{PhysRevA.80.013608}
M.~Th\o{}gersen, D.~V. Fedorov, A.~S. Jensen, B.~D. Esry, and Y.~Wang.
\newblock Conditions for efimov physics for finite-range potentials.
\newblock {\em Phys. Rev. A}, 80:013608, Jul 2009.

\bibitem{PhysRevA.78.020501}
M.~Th\o{}gersen, D.~V. Fedorov, and A.~S. Jensen.
\newblock Universal properties of efimov physics beyond the scattering length
  approximation.
\newblock {\em Phys. Rev. A}, 78:020501, Aug 2008.

\end{thebibliography}

\end{document}